\documentstyle[aps]{revtex}

\newcommand{\beq}{\begin{equation}}
\newcommand{\eeq}{\end{equation}}
\newcommand{\four} {  {}^{(4)}\kern-1pt  }
\newcommand{\ben}{\begin{eqnarray}}
\newcommand{\een}{\end{eqnarray}}

\catcode`@=11     
\def\eqalign#1{\null\,\vcenter{\openup\jot\m@th
  \ialign{\strut\hfil$\displaystyle{##}$&$\displaystyle{{}##}$\hfil
      \crcr#1\crcr}}\,}
\catcode`@=12   

\begin{document}

\title{
$~$
{\Large {\bf Phenomenology of
Planck-scale Lorentz-symmetry test theories}}}

\author{{\bf Giovanni~AMELINO-CAMELIA}}
\address{Dipart.~Fisica,
Univ.~Roma ``La Sapienza'', and INFN Sez.~Roma1\\
P.le Moro 2, 00185 Roma, Italy}

\maketitle

\begin{abstract}
In the recent quantum-gravity literature there has been
strong interest in the possibility
of Planck-scale departures from Lorentz symmetry,
including possible modifications of the
energy/momentum dispersion relation.
I stress that a meaningful characterization
of the progress of experimental bounds on these Planck-scale
effects requires the analysis of some reference test theories,
and I propose to focus on
two ``minimal'' test theories, a pure-kinematics test theory
and an effective-field-theory-based test theory.
I illustrate some features of the phenomenology based on these
test theories considering some popular strategies for constraining
Planck-scale effects, and in particular I observe
that sensitivities that are already in the Planck-scale range
for some parameters of the two test theories can be achieved
using observations of TeV photons from Blazars,
both using the so-called ``gamma-ray time-of-flight analyses''
and using the now robust evidence
of absorption of TeV photons.
Instead the Crab-nebula
synchrotron-radiation analyses, whose preliminary sensitivity estimates
raised high hopes,
actually do not lead to any bound on the parameters of
the two ``minimal'' test theories.
The Crab-nebula synchrotron-radiation analyses do however
constrain some possible generalizations of one
of the minimal test theories.
As an example of forthcoming data which could provide
extremely stringent (beyond-Planckian) limits
on the two minimal test theories I consider the possibility
of studies of the GZK cutoff for cosmic-rays.

\end{abstract}

\bigskip
\bigskip

\section{Different perspectives on the fate of Lorentz symmetry
in quantum spacetime}
The fact that Lorentz symmetry is such a crucial ingredient
of our present description
of the fundamental laws of physics has motivated a large effort
to test this symmetry to the highest possible precision.
In addition to the general interest in probing the robustness
of the principles we hold as fundamental,
recently tests of Lorentz symmetry have attracted
interest also as a result of the realization that in various
approaches to the quantum-gravity problem one encounters nonclassical
features of spacetime that lead to small departures from
Lorentz symmetry. A quantum-gravity-motivated
phenomenology of departures from Lorentz symmetry was proposed
in Ref.~\cite{grbgac}. The idea
that Lorentz symmetry might be only an approximate symmetry
has then been considered
in quantum-gravity models based on spacetime foam
pictures~\cite{garayPRL}, in loop quantum gravity
models~\cite{gampul,mexweave}
and in noncommutative geometry
models~\cite{gacmajid,susskind,dougnekr,gacdsr,dsrmost1},
including some scenarios for noncommutative geometry
that are relevant in string theory~\cite{susskind,dougnekr}.

At a strictly phenomenological level one can view
this interest in possible Planck-scale departures from Lorentz
symmetry as originating from the idea that the sought quantum gravity
might involve some sort of ``granularity" of spacetime (``spacetime quanta"),
and on the basis of experience with certain physical systems
(especially condensed-matter systems)
one can expect that granularity of the medium in which propagation occurs
might lead to energy-dependent corrections~\cite{grbgac} to the dispersion
relation. At energies much larger than the particle mass but smaller
than the granularity (Plankian) energy scale,
the dispersion relation could be of the type\footnote{In the literature
 the correction term is
 treated equivalently as a $\vec{p}^2 E^n$ correction and as
 a $E^{n+2}$ correction, since one is anyway only interested in
 leading-order corrections in processes involving high-energy
 ($\vec{p}^2 \simeq E^2$)
 particles. Of course, the symbol $m$ is meaningful as
 the rest energy of the particle (a low-energy concept) only
 if the correction term vanishes for particles at rest (the
 case of a $\vec{p}^2 E^n$ correction).}
 \begin{equation}
 m^2 \simeq E^2 - \vec{p}^2
 +  \eta \vec{p}^2 \left({E^n \over E^n_{p}}\right)
 + O({E^{n+3} \over E^{n+1}_{p}})
 ~
 \label{displead}
 \end{equation}
where $E_p \simeq 1.2 {\cdot} 10^{16} TeV$ is the Planck scale,
$\eta$ parametrizes the ratio between the Planck scale and
the scale of quantization of spacetime,
and the power $n$ is a key
characteristic of the magnitude of the effects
to be expected.

Of course, different intuitions for the right path toward
quantum gravity may lead to
different expectations with respect to this possible departures
from Lorentz symmetry.
And it is useful to realize
that each quantum-gravity research line
can be connected with one of
three perspectives on the problem: the particle-physics perspective,
the general-relativity perspective and the condensed-matter perspective.

From a particle-physics perspective it is natural to attempt to
reproduce as much as possible the successes of the Standard Model
of particle physics.
One is tempted to see gravity simply as one more gauge interaction.
From this particle-physics perspective a
natural solution of the quantum gravity problem would be String-Theory-like:
a quantum gravity whose core features
are essentially described in terms of graviton-like exchange
in a background classical spacetime.
And from this particle-physics perspective there is clearly no in-principle
reason to renounce to exact Lorentz symmetry, at least
as long as
Minkowski classical spacetime is an admissible background spacetime.
Still, a breakup of Lorentz symmetry,
in the sense of spontaneous symmetry breaking,
is of course possible.
And this possibility has been studied extensively~\cite{susskind,dougnekr}
over the last few years, particularly in String Theory, which is
the most mature quantum-gravity approach that emerged from the particle-physics
perspective.

The general-relativity perspective
naturally leads to reject the use of a background spacetime, and
this is widely acknowledged~\cite{crLIVING,ashtNEW,leeLQGrev,thieREV}.
Although less publicized, there is also growing awareness
of the fact that the development of general relativity relied heavily
on the careful consideration of the in-principle limitations that
measurement procedures can encounter. Think for example of the limitations
that the speed-of-light limit imposes on certain setups for clock synchronization.
In light of the various arguments
suggesting that, whenever both quantum mechanics and general relativity
are taken into account, there should be an in-principle
limitation to the localization of a spacetime
point (an event),
the general-relativity perspective
invites one to renounce to any direct reference
to a classical spacetime~\cite{dopl1994,ahlu1994,ng1994,gacmpla}.
Indeed this requirement
that the in-principle measurability limitations be reflected by the adoption
of a corresponding measurability-limited description of spacetime,
is another element of intuition which is guiding quantum-gravity
research from the general-relativity perspective.
This naturally leads one to consider certain types of
discretized spacetimes, as in the Loop
Quantum Gravity approach\cite{crLIVING,ashtNEW,leeLQGrev,thieREV},
or noncommutative spacetimes~\cite{dopl1994,ahlu1994}.
Results obtained over the last few years
indicate that
from this general-relativity perspective
some Planck-scale departures from Lorentz symmetry are
naturally expected (although not automatic).
Loop quantum gravity and other discretized-spacetime quantum-gravity
approaches appear to require~\cite{gampul,mexweave,kodadsr} some departures,
governed by the Planck scale, from the familiar
(continuous) Lorentz symmetry.
And Planck-scale departures
from Lorentz symmetry might be inevitable
in noncommutative spacetimes, as shown in
several recent
studies~\cite{susskind,dougnekr,gacdsr,dsrmost1,majrue,kpoinap}.

The third possibility is a condensed-matter perspective
(see, {\it e.g.}, the research programs of Refs.~\cite{volovik}
and \cite{laugh})
on the quantum-gravity problem,
in which some of the familiar properties of spacetime are
only emergent.
Condensed-matter theorists are used to describe some of
the degrees of freedom that are measured in the laboratory
as collective excitations within
a theoretical framework whose primary description is given
in terms of much different, and often practically unaccessible,
fundamental degrees of freedom.
Close to a critical point some symmetries arise for
the collective-excitations theory, but these symmetries
do not carry the significance of fundamental symmetries,
and are in fact lost as soon as the theory is probed somewhat
away from the critical point. Notably,
some familiar systems are known to exhibit special-relativistic invariance
in certain limits, even though, at a more fundamental level,
they are described in terms of a nonrelativistic theory.
Clearly from this (relatively new) condensed-matter perspective
on the quantum-gravity problem it is natural to see the familiar
classical continuous Lorentz symmetry only as an approximate
(emergent) symmetry.

The interest in testing the idea of Planck-scale departures from
Lorentz symmetry is also due to the differences between these
alternative perspectives on the quantum-gravity problem.
Any experimental hint on the fate of Lorentz symmetry at the Planck
scale might allow us to establish which (if any) of these
perspectives on the quantum-gravity problem is to be favoured.

\section{Planck-scale departures from Lorentz symmetry,
modified dispersion relations and test theories}
Motivation for the study of Planck-scale departures from Lorentz symmetry
does not simply come from the mentioned general perspectives
on the quantum-gravity problem:
for some of the theoretical frameworks that are being considered
in quantum-gravity research evidence of such
 departures from Lorentz symmetry has been found.
In this section I
start with a brief description of how modified dispersion
relations arise\footnote{I discuss noncommutative spacetimes
and the Loop Quantum Gravity approach, which are the
best understood Planck-scale frameworks in which
it appears that the dispersion relation is Planck-scale modified.
But other types of intuitions about the quantum-gravity problem
may lead to modified dispersion relations, including some realizations
of the idea of ``spacetime foam''~\cite{grbgac,garayPRL,emn},
which allow an analogy with the laws of particle
propagation in a thermal environment~\cite{grbgac,garayPRL,gacpi}.}
in the study of noncommutative spacetimes
and in the study of loop quantum gravity.

These results should provide guidance in setting up a phenomenology
for the Planck-scale departures from Lorentz symmetry.
In particular I want to stress that it is necessary
for this phenomenology
to rely on some reference test theories,
which should be inspired by the results obtained in
the study of noncommutative spacetimes
and in the study of loop quantum gravity.
In the second part of this section
I discuss two such test theories on which
it might be appropriate to focus.

\subsection{Modified dispersion relations in
canonical noncommutative spacetime}
The noncommutative spacetimes in which modifications of the dispersion
relation are being most actively considered all fall within
the following rather general parametrization of
noncommutativity of the spacetime coordinates:
\begin{equation}
\left[x_\mu,x_\nu\right] = i \theta_{\mu \nu}
+ i \rho^\beta_{\mu \nu} x_\beta ~.
\label{all}
\end{equation}
It is convenient to first focus on the special case $\rho = 0$,
the ``canonical noncommutative spacetimes''
\begin{equation}
\left[x_\mu,x_\nu\right] = i \theta_{\mu \nu}
~.
\label{cano}
\end{equation}

Of course, the natural first guess for introducing dynamics
in these spacetimes is a quantum field theory formalism.
And indeed, for the special case $\rho = 0$,
an approach to the
construction of a quantum field theory has been developed rather
extensively~\cite{susskind,dougnekr}.
While most aspects of these field theories closely resemble their
commutative-spacetime counterparts, a surprising feature
that emerges is the
so-called ``IR/UV mixing"~\cite{susskind,dougnekr,gianlucaken}:
the high-energy sector of the theory does not decouple
from the low-energy sector.
Connected with this IR/UV mixing is the type of modified dispersion
relations that one encounters in field theory on canonical noncommutative
spacetime, which in general take the form
\begin{equation}
 m^2 \simeq E^2 - \vec{p}^2
+ \frac{\alpha_1}{p^\mu \theta_{\mu \nu} \theta^{\nu \sigma} p_\sigma}
+ \alpha_2
m^2 \ln \left( p^\mu \theta_{\mu \nu} \theta^{\nu \sigma} p_\sigma \right)
+ \dots
~,
\label{dispCANO}
\end{equation}
where the $\alpha_i$ are parameters, possibly taking different
values for different particles (the dispersion relation is not ``universal''),
that depend on various aspects
of the field theory, including its field content and the nature
of its interactions.
The fact that this dispersion relation can be singular in the infrared
is a result of the IR/UV mixing. A part of the infrared singularity
could be removed by introducing (exact) supersymmetry, which typically
leads to $\alpha_1 = 0$.

The implications
of this IR/UV mixing for dynamics are still not fully understood,
and there is still justifiable skepticism~\cite{skepticNCFT}
toward the reliability of
the type of field-theory construction adopted so far.
I think it is legitimate to even wonder whether
a field-theoretic formulation of the dynamics is at all truly compatible
with the canonical spacetime noncommutativity.
The Wilson decoupling between IR and UV degrees of freedom
is a crucial ingredient of most applications of field
theory in physics, and it is probably incompatible with
canonical noncommutativity: the associated uncertainty
principle of the type $\Delta x_\mu \Delta x_\nu \ge \theta_{\mu \nu}$
implies that it is not possible to probe short distances
(small, say, $\Delta x_1$) without probing simultaneously
the large-distance regime ($\Delta x_2 \ge \theta_{2,1 }/\Delta x_1$).

In any case, the presence of modified dispersion relations
in canonical noncommutative spacetime should be expected, since
Lorentz symmetry is ``broken'' by the tensor $\theta_{\mu \nu}$.
An intuitive characterization of this Lorentz-symmetry breaking
can be obtained by looking at wave exponentials. The Fourier
theory in canonical noncommutative spacetime is based~\cite{wessLANGUAGE}
on simple wave exponentials $e^{i p^\mu x_\mu}$ and from
the $[x_\mu,x_\nu] = i \theta_{\mu \nu}$
noncommutativity relations one finds that
\begin{equation}
e^{i p^\mu x_\mu} e^{i k^\nu x_\nu}
= e^{-\frac{i}{2} p^\mu
\theta_{\mu \nu} k^\nu} e^{i (p+k)^\mu x_\mu} ~,
\label{expprodcano}
\end{equation}
{\it i.e.} the Fourier parameters $p_\mu$ and $k_\mu$ combine just as
usual, but there is the new ingredient of the overall $\theta$-dependent
phase factor.

The fact that momenta combine in the usual way reflects the fact that
the transformation rules for energy-momentum from one
(inertial) observer to another are still the familiar, undeformed,
Lorentz transformation rules. However, the product of wave exponentials
depends on $p^\mu \theta_{\mu \nu} k^\nu$; it depends on the ``orientation"
of the energy-momentum vectors $p^\mu$ and $k^\nu$
with respect to the $\theta_{\mu \nu}$ tensor.
The $\theta_{\mu \nu}$ tensor plays the role of a
background that
identifies a preferred class of inertial observers\footnote{Note that
these remarks apply to canonical noncommutative spacetimes
as studied in the most recent (often String-Theory inspired) literature,
in which $\theta_{\mu \nu}$ is indeed simply a tensor (for a given
observer, an antisymmetric matrix of numbers).
I should stress however that the earliest studies of canonical noncommutative
spacetimes (see Ref.~\cite{dopl1994} and follow-up work)
considered a $\theta_{\mu \nu}$ with richer mathematical properties,
notably with nontrivial algebra relations with the spacetime coordinates.
In that earlier, and more ambitious, setup it is not obvious that Lorentz
symmetry would be broken: the fate of Lorentz symmetry
may depend on the properties attributed to $\theta_{\mu \nu}$.}.
Different particles can be affected by the presence of this background
in different ways, leading to the emergence of different
dispersion relations.
All this is consistent with indications of the mentioned
popular field theories in canonical noncommutative spacetimes.

\subsection{Modified dispersion relations in
 $\kappa$-Minkowski noncommutative spacetime}
In canonical noncommutative spacetimes Lorentz symmetry is ``broken''
and there is growing evidence that Lorentz symmetry breaking occurs
for most choices of the tensors $\theta$ and $\rho$.
It is at this point clear, in light of several recent results,
that the only way to preserve Lorentz symmetry
is the choice $\theta = 0 =\rho $, {\it i.e.} the case in which
there is no noncommutativity
and one is back to the familiar classical
commutative Minkowski spacetime.
When noncommutativity is present
Lorentz symmetry is usually
broken, but
recent results suggest that for some special choices of the
tensors $\theta$ and $\rho$
Lorentz symmetry might be deformed,
in the sense of the recently proposed ``doubly-special relativity''
scenario~\cite{gacdsr}, rather than broken.
In particular, this appears to be the case for the Lie-algebra
$\kappa$-Minkowski~\cite{majrue,kpoinap,gacmajid,lukieFT,gacmich,wesskappa}
noncommutative spacetime ($l,m = 1,2,3$)
\begin{equation}
\left[x_m,t\right] = {i \over \kappa} x_m ~,~~~~\left[x_m, x_l\right] = 0 ~.
\label{kmindef}
\end{equation}

$\kappa$-Minkowski
is a Lie-algebra spacetime that clearly enjoys classical space-rotation
symmetry; moreover, at least in a Hopf-algebra sense (see, {\it e.g.},
Ref.~\cite{gacmich}), $\kappa$-Minkowski
is invariant under ``noncommutative translations''.
Since I am focusing here on Lorentz symmetry,
it is particularly noteworthy that in $\kappa$-Minkowski
boost transformations are necessarily modified~\cite{gacmich}.
A first hint of this comes from the necessity of a deformed
law of composition of momenta, encoded
in the so-called coproduct (a standard structure for a Hopf algebra).
One can see this clearly by considering the Fourier tranform.
It turns out~\cite{gacmajid,lukieFT} that in
the $\kappa$-Minkowski case the correct formulation of the Fourier theory
requires a suitable ordering prescription
for wave exponentials. From
\begin{equation}
 :e^{i k^\mu x_\mu}: \equiv e^{i k^m x_m} e^{i k^0 x_0}
~,
\label{order}
\end{equation}
as a result of $[x_m,t] = i x_m/\kappa$
(and $[x_m, x_l] = 0$),
it follows that
the wave exponentials combine in a nontrivial way:
\begin{equation}
(:e^{i p^\mu x_\mu}:) (:e^{i k^\nu x_\nu}:) =
:e^{i (p \dot{+} k)^\mu x_\mu}:
\quad.
\label{expprodlie}
\end{equation}
The notation ``$\dot{+}$" here introduced reflects the
behaviour of the mentioned ``coproduct"
composition of momenta:
\begin{equation}
p_\mu \dot{+} k_\mu \equiv \delta_{\mu,0}(p_0+k_0) + (1-\delta_{\mu,0})
(p_\mu +e^{\lambda p_0} k_\mu) ~. \label{coprod}
\end{equation}

As argued in Refs.~\cite{gacdsr} the nonlinearity of the law of composition
of momenta might require an absolute (observer-independent) momentum scale,
just like upon introducing a nonlinear law of composition of velocities
one must introduce the absolute observer-independent scale of
velocity $c$. The inverse of the noncommutativity scale $\lambda$
should play the role of this absolute momentum scale.
This invites one to consider the possibility~\cite{gacdsr}
that the transformation laws for energy-momentum
between different observers would have two invariants, $c$ and $\lambda$,
as required in ``doubly-special relativity''~\cite{gacdsr}.

On the basis of (\ref{coprod}) one is led~\cite{majrue,kpoinap,gacmajid}
to the following result for the form of the energy/momentum
dispersion relation
\begin{equation}
\left(\frac{2}{\lambda}\sinh\frac{\lambda m}{2}\right)^2 =
\left(\frac{2}{\lambda}\sinh\frac{\lambda E}{2}\right)^2-
e^{\lambda E}\vec{p}^2
~,
\label{dispkpoin}
\end{equation}
which for low momenta takes the approximate form
\begin{equation}
m^2 \simeq E^2 - \vec{p}^2
- \lambda E \vec{p}^2
~.
\label{dispkpoinlimit}
\end{equation}
Actually, the precise form of the dispersion relation
may depend on the choice of ordering prescription
for wave exponentials~\cite{gacmich}
((\ref{dispkpoin}) follows form (\ref{order})),
and this point deserves further scrutiny.
But even setting aside this annoying ordering ambiguity,
there appear to be severe obstructions~\cite{lukieFT,gacmich} for
a satisfactory formulation of a
quantum field theory in $\kappa$-Minkowski.
The techniques that were rather straightforwardly applied
for the construction of field theory in canonical noncommutative spacetime
do not appear~\cite{lukieFT,gacmich}
to be applicable in the $\kappa$-Minkowski case.
It is not unplausible that the ``virulent''  $\kappa$-Minkowski
noncommutativity may require
some departures from a standard field-theoretic setup.

\subsection{Modified dispersion relation in Loop Quantum Gravity}
Loop Quantum Gravity is one of the most ambitious
approaches to the quantum-gravity
problem, and its
understanding is still in a relatively early stage.
As presently understood, Loop Quantum Gravity predicts an inherently
discretized spacetime~\cite{crLIVING,ashtNEW,leeLQGrev},
and this occurs in a rather compelling way: it is not that one introduces
by hand an {\it a priori} discrete background spacetime; it is rather
a case in which a background-independent analysis ultimately
leads, by a sort of self-consistency, to the emergence of
discretization.
There has been much discussion recently, prompted by the
studies~\cite{grbgac,gampul,mexweave},
of the possibility that this discretization
might lead to broken Lorentz symmetry and a modified dispersion relation.
Although there are
cases in which a discretization is compatible
with the presence of continuous classical
symmetries~\cite{simonecarlo,areanew},
it is of course natural, when adopting a discretized spacetime,
to put Lorentz symmetry under careful scrutiny.
Arguments presented in Refs.~\cite{gampul,mexweave}
suggest that Lorentz
symmetry might indeed be broken in Loop Quantum Gravity.

Moreover, very recently Smolin, Starodubtsev and I proposed~\cite{kodadsr}
(also see the follow-up study in Ref.~\cite{jurekkodadsr})
a mechanism such that Loop Quantum Gravity
would be described at the most fundamental level as a theory that in the
flat-spacetime limit admits deformed Lorentz symmetry,
in the sense of the ``doubly-special relativity''
scenario~\cite{gacdsr}.
Our argument originates from the role that certain quantum symmetry groups
(``q-deformed algebras'')
have in the Loop-Quantum-Gravity description of spacetime with
a cosmological constant, and observing that in the flat-spacetime limit
(the limit of vanishing cosmological constant)
these quantum groups might not contract to a classical Lie algebra,
but rather contract to a quantum (Hopf) algebra.

All these studies point to the presence of a modified dispersion
relation, although different arguments lead to different
intuition for the form of the dispersion relation.
A definite result might have to wait for the solution of
the well-known ``classical-limit problem" of
Loop Quantum Gravity. We are presently
unable to recover from this full quantum-gravity theory the limiting
case in which the familiar quantum-field-theory description of
particle-physics processes in a classical background spacetime applies.
Some recent results~\cite{varadaADD,ashteADD},
which tackle the problem of reproducing
Fock-space quantization from the Loop-Quantum-Gravity framework,
may provide the first ingredients of such a formulation.
But other recent studies appear to suggest\cite{gampulDENSI}
that in the same contexts in which departures from Lorentz
symmetry may be revealed one should adopt a density-matrix formalism,
and then the whole picture would collapse to the familiar Lorentz-invariant
quantum-field-theory description in contexts involving both relatively
low energies and relatively low boosts with respect to the center-of-mass
frame (e.g. the particle-physics collisions studied at several
particle accelerators).

\subsection{Some issues relevant for the proposal of test theories}
While the first few years
of work on this idea of Planck-scale departures from Lorentz
symmetry
were necessarily based on rather preliminary analyses,
with the only objective of establishing the point that
Planck-scale sensitivity could be achieved in some cases,
I want to stress that
we should now gear up for a more ``mature'' phase of work
on quantum-gravity phenomenology, in which the development
and analysis of some carefully crafted test theories takes center stage.
The results I briefly summarized in the previous three subsections
suggest that in the analysis of noncommutative spacetimes
and in the analysis of Loop Quantum Gravity, the two approaches that
provide most of the motivation for this phenomenology,
we are getting closer to obtaining truly characteristic predictions,
prediction that could be used to falsify the corresponding
theoretical scheme. But there are
a few open issues which do not at present allow us
to describe in detail a falsifiable prediction,
and therefore, for now, the phenomenology must rely
on some appropriately structured test theories.
These test theories should on the one hand
reflect the points we do understand of
these quantum-gravity approaches and on the other hand
they should limit as much as possible
the risk of assuming properties that could turn out not
to be verified once we understand the formalisms better.

The test theories should also be used for bridging the gap
between the experimental data and the
analysis of the formalisms.
The test theories should provide a common language in assessing
the progresses made in improving the sensitivity of experiments,
a language that must also be suitable for access from the side of
those working at the development of the quantum-gravity/quantum-spacetime
theories.

As we contemplate the challenge of developing such carefully-balanced
test theories it is important to observe that
the most robust part of the results I summarized in the previous
three subsections
is clearly the emergence of modified dispersion relations.
Therefore if one could set up experiments testing
directly the dispersion relation
the resulting limits would have wide applicability.
In principle one could investigate the form of the
dispersion relation directly by making simultaneous
measurements of energy and space-momentum;
however, it is easy to see that achieving Planck-scale
sensitivity in such a direct test is well beyond our capabilities.

Useful test theories on which to base the relevant phenomenology
must therefore combine
the ingredient of the dispersion relation with other ingredients.
As I shall discuss in greater detail later in this section,
there are three key issues for this test-theory development:

\begin{list}{}{}

\item (i) in presence of the modified dispersion relation should
we still assume the validity
of the relation $v = dE/dp$ between the speed of a particle
and its dispersion relation? (here $dE/dp$ is the derivative of
the function $E(p)$ which of course is implicitly introduced
through the dispersion relation)

\smallskip

\item (ii) in presence of the modified dispersion relation should
we still assume the validity of the standard law of
energy-momentum conservation?

\smallskip

\item (iii) in presence of the modified dispersion relation
which formalism should be adopted for the description
of dynamics?

\end{list}


The fact that these are key issues is also a consequence
of the type of data that we expect to have access to,
as I shall discuss here later.

Unfortunately on these three key points the quantum-spacetime
pictures which are providing motivation for the study
of Planck-scale modifications of the dispersion relation,
reviewed in the previous three subsections,
are not providing much guidance yet.

For example, in Loop Quantum Gravity,
while we do have evidence that the dispersion relation should
be modified, we do not yet have a clear indication concerning
whether the law of energy-momentum conservation should also
be modified and we also cannot yet robustly establish whether
the relation $v=dE/dp$ should be preserved.
Moreover, the ``classical-limit problem'',
as mentioned, also affects the choice of
formalism to be adopted for the description
of dynamics.
It is not at all clear how and in which regimes a field-theoretic
setup should be available,
and
some recent studies appear to suggest~\cite{gampulDENSI}
that in the same contexts in which departures from Lorentz
symmetry may be revealed one should also adopt a density-matrix formalism.
We should therefore be prepared for surprises in the description
of dynamics.

Similarly in the analysis of noncommutative
spacetimes we are close to establishing in
rather general terms that some modification of the dispersion relation
is inevitable,
but other aspects of the framework have not yet been clarified.
While most of the literature for canonical noncommutative spacetimes
assumes~\cite{susskind,dougnekr}
that the law of energy-momentum conservation should not be modified,
most of the literature for $\kappa$-Minkowski spacetime
argues in favour of a modification
(perhaps consistent with the corresponding
doubly-special-relativity criteria~\cite{gacdsr})
of the law of energy-momentum conservation.
There is also still no consensus
on the relation between speed and dispersion relation,
and particularly in the $\kappa$-Minkowski literature
some departures from the $v=dE/dp$ relation are actively
considered~\cite{Kosinski:2002gu,Mignemi:2003ab,Daszkiewicz:2003yr,jurekREV}.
And concerning the formalism to be used for the description
of dynamics in a noncommutative spacetime, while everybody's first
guess is the field-theoretic formalism, the fact that
attempts at a field theory formulation encounter so many difficulties
(the IR/UV mixing for the canonical-noncommutative spacetime case
and the even more pervasive shortcomings of the proposals for a
field theory in $\kappa$-Minkowski)
must invite one to contemplate possible alternative formulations
of dynamics.

Clearly the situation on the theory side invites us to be prudent:
if a given phenomenological picture relies on too many assumptions
on Planck-scale physics it is likely that it might not reproduce
any of the mentioned quantum-gravity and/or quantum-spacetime models
(when these models are eventually fully understood they will
give us their own mix of Planck-scale features, which is difficult to
guess at the present time).
On the other hand it is necessary for the robust development
of a phenomenology to adopt well-defined test theories.
Without reference to a well-balanced set of test theories
it is impossible to compare the limits obtained in different
experimental contexts, since each experimental context may require
different ``ingredients" of Planck-scale physics.
And it is of course meaningless to compare limits obtained
on the basis of
different conjectures for the Planck-scale regime.

\subsection{A test theory for pure kinematics}
The majority (see, {\it e.g.},
Refs.~\cite{billetal,kifu,ita,aus,gactp})
of studies concerning Planck-scale modifications of the
dispersion relation adopt the phenomenological formula
\begin{equation}
 m^2 \simeq E^2 - \vec{p}^2
+  \eta \vec{p}^2 \left({E^n \over E^n_{p}}\right)
+ O({E^{n+3} \over E^{n+1}_{p}})
~,
\label{displeadbis}
\end{equation}
with real $\eta$ of order $1$ and integer $n$.
This formula is compatible with some
of the results obtained in the Loop-Quantum-Gravity approach
and reflects the results obtained in $\kappa$-Minkowski and
other noncommutative spacetimes (but, as mentioned, in the special case
of canonical noncommutative spacetimes one encounters
a different, infrared singular, dispersion relation).

As mentioned, on the basis of the status on the theory side,
a prudent approach in combining the dispersion relation
with other ingredients is to be favoured.
Since basically all experimental situations will involve
some aspects of kinematics that go beyond the dispersion
relation (while there are some cases in which the dynamics,
the interactions among particles, does not play a role),
and taking into account the mentioned difficulties in establishing
what is the correct formalism for the description
of dynamics\footnote{I am here using the expression ``dynamics at
the Planck scale'' with some license.
Of course, in our phenomenology we will not be sensitive directly
to the dynamics at the Planck scale. However, as I discuss
in greater detail in the next subsection, if the arguments
that encourage the use of new descriptions of dynamics at
the Planck scale are correct, then a sort of ``order of limits problem''
clearly arises. Our experiments will involve energies much lower
than the Planck scale, and we know that in the infrared limit
the familiar formalism with field-theoretic description of
dynamics and Lorentz invariance will hold. So we would need to
establish whether experiments that are sensitive to Planck-scale
departures from Lorentz symmetry could also be sensitive to
Planck-scale departures from the field-theoretic description
of dynamics. Since we still know very little about this
alternative descriptions of dynamics a prudent approach,
avoiding any assumption about the description of dynamics
is certainly preferable.}
at the Planck scale,
most authors prefer to prudently combine the dispersion relation
with other ``purely kinematical" aspects of Planck-scale physics.

Already in the first studies~\cite{grbgac} that proposed a phenomenology
based on (\ref{displeadbis}) it was assumed
that the dispersion relation would still be ``universal''
(same for all particles)
and that even at the Planck scale the familiar
description of ``group velocity", obtained from the dispersion relation
according to $v=dE/dp$, should hold\footnote{As mentioned,
this assumption is not guaranteed to apply to the formalisms
of interest, and indeed several authors
have considered
alternatives~\cite{Kosinski:2002gu,Mignemi:2003ab,Daszkiewicz:2003yr,jurekREV}.
While the studies advocating
alternatives to $v = dE/dp$ rely of a large variety of
arguments (some more justifiable some less),
in my own perception~\cite{gianluFranc}
a key issue here is whether quantum gravity leads to a modified
Heisenberg uncertainty principle, $ [x,p] =1 + F(p)$. Assuming a Hamiltonian
description is still available, $v = dx/dt \sim [x,H(p)]$,
the relation $v = dE/dp$ essentially follows from $ [x,p] =1$.
But if $ [x,p] \neq 1$ then  $v = dx/dt \sim [x,H(p)]$ would not lead
to $v = dE/dp$. And there is much discussion in the quantum-gravity
community of the possibility of
modifications of the Heisenberg uncertainty principle at the Planck scale.}.

In other works motivated by the analysis reported in Ref.~\cite{grbgac}
another key kinematical feature was introduced:
starting with the studies reported in Refs.~\cite{kifu,ita,aus,gactp}
the dispersion relation (\ref{displeadbis}) and
the velocity relation $v = dE/dp$
were combined with the assumption that the law of energy-momentum conservation
should not be modified at the Planck scale, so that, for example,
in a $a + b \rightarrow c + d$ particle-physics process one would have
\begin{equation}
E_a + E_b = E_c + E_d
~,
\label{econs}
\end{equation}
\begin{equation}
\vec{p}_a + \vec{p}_b = \vec{p}_c + \vec{p}_d
~.
\label{pcons}
\end{equation}


The elements I described in this Section compose a
quantum-gravity phenomenology test theory
that has already been widely considered in
the literature, even though it was never previously
characterized in detail.
In the following I will refer to this test theory as the ``minimal
AEMNS test
theory"\footnote{I am using ``AEMNS'' on the basis of the initials of the
names of the authors in Ref.\cite{grbgac}, which first proposed a
phenomenology based on the dispersion relation (\ref{displeadbis}).
But as mentioned the full test theory, as presently used in most
studies, only emerged gradually in follow-up work. In particular,
there was no discussion of energy-momentum conservation
in Ref.~\cite{grbgac}. Unmodified energy-momentum conservation
was introduced in Refs.~\cite{kifu,ita,aus,gactp}.
Concerning this  ``minimal AEMNS test
theory" it should also be noticed that Ellis, Mavromatos and
Nanopoulos actually favour~\cite{nycksync} a quantum-gravity
approach in which
the modification of the dispersion relation is not universal,
and therefore would not fit within the confines of
the ``minimal AEMNS test
theory" (although, of course, nonuniversality can be accommodated in
a straightforward generalization of the test theory).},
and I will assume that experimental
bounds on this test theory should be placed by using only
the following assumptions:

(minAEMNS.1) the dispersion relation is of the form
\begin{equation}
 m^2 \simeq E^2 - \vec{p}^2
+  \eta \vec{p}^2 \left({E^{n} \over E^{n}_{p}}\right)
+ O({E^{n+3} \over E^{n+1}_{p}})
~,
\label{displeadNONUNImin}
\end{equation}
where $\eta$ and $n$ are universal (same value for every particle
and for both helicities/polarizations of a given particle);

(minAEMNS.2) the velocity of a particle can be obtained from
the dispersion relation using $v=dE/dp$;

(minAEMNS.3) the law of energy-momentum conservation is not modified;

(minAEMNS.4) nothing is assumed about dynamics ({\it i.e.}
the analysis of this test theory will be limited to classes of
experimental data that involve pure kinematics, without any
role for dynamics).

While this ``minimal'' version of the test theory appears
to deserve to be
the primary focus of AEMNS-based
phenomenology work, it is of course legitimate
to consider some possible generalizations,
including a nonuniversality of the effects (allowing for
different values of the dispersion-relation-modification
parameters for different particles).

\subsection{A test theory based on low-energy effective field theory}
The AEMNS test theory has the merit of relying only on a relatively
small network of assumptions on kinematics, without assuming
anything about the role of the Planck scale in dynamics.
However, of course, this justifiable prudence turns into
a severe limitation on the class of experimental contexts
which can be used to constrain the parameters of the test theory.
It is in fact rather rare that a phenomenological analysis
can be completed without using
any aspects of the interactions among the particles involved
in the relevant processes.
The desire to be able to analyze a wider class of experimental contexts
is therefore providing motivation for the development of
test theories more ambitious than the AEMNS test theory,
with at least some elements of dynamics.
This is understandable but, in light of the situation on the theory
side, work with one of these more ambitious test theories
should proceed with the awareness that there is a high risk that
it may turn out that
none of the quantum-gravity approaches which are being pursued
is reflected in the test theory.

One reasonable possibility to consider, when the urge to analyze data
that involve some contamination from dynamics cannot be resisted,
is the one of describing dynamics within the framework
of low-energy effective field theory.
In this subsection I want to discuss a test theory
which is indeed based on low-energy effective field theory,
and has emerged from the work recently reported in
Ref.~\cite{rob} (which is rooted in part in the earlier Ref.~\cite{gampul}).

Before a full characterization of this test theory
I should first warn the reader that there might be some
severe limitations for the applicability of
low-energy effective field theory to the investigation of
Planck-scale physics,
especially when departures from Lorentz symmetry are present.

A significant portion of the quantum-gravity community
is in general, justifiably, skeptical about the results obtained
using low-energy effective field theory in
analyses relevant for the quantum-gravity problem.
After all the first natural prediction of
low-energy effective field theory
in the gravitational realm is a value of the energy density
which is some 120 orders of magnitude greater
than allowed by observations\footnote{And the
outlook of low-energy effective field theory
in the gravitational realm does not improve much through the observation
that exact supersymmetry could protect from the emergence of any energy density.
In fact, Nature clearly does not have supersymmetry at least
up to the TeV scale, and this would still lead to a natural prediction
of the cosmological constant which is some 60 orders of magnitude too high.}.

As a result of the different perspectives on the quantum-gravity
problem, which I already described in Section~1,
there are on the one hand numerous
researchers who are skeptical about any results obtained
using low-energy effective field theory in
analyses relevant for the quantum-gravity problem,
but there are on the other hand
also quite a few researchers
interested in the quantum-gravity problem
who are completely serene in assuming
that all quantum-gravity effects
should be describable in terms of effective field theory
in low-energy situations.

I feel that, while of course an effective-field-theory
description may well turn out to be correct in the end,
the {\it a priori} assumption
that a description in terms of
effective low-energy field-theory should work is rather naive.
If the arguments
that encourage the use of new descriptions of dynamics at
the Planck scale are correct, then a sort of ``order of limits problem''
clearly arises. Our experiments will involve energies much lower
than the Planck scale, and we know that in some limit (a limit
that characterizes our most familiar observations)
the field-theoretic description
and Lorentz invariance will hold. So we would need to
establish whether experiments that are sensitive to Planck-scale
departures from Lorentz symmetry could also be sensitive to
Planck-scale departures from the field-theoretic description
of dynamics.
As an example, let me mention the possibility (not unlikely in
a context which is questioning the fate of Lorentz symmetry)
that quantum gravity would admit a field-theory-type description
only in reference frames in which the process of interest
is essentially occurring in its center of mass
(no ``Planck-large boost''~\cite{mg10qg1}
with respect to center-of-mass frame).
The field theoretic description could emerge in
a sort of ``low-boost limit'', rather than the expected
low-energy limit.
The regime of low boosts with respect the center-of-mass frame
is often indistinguishable with respect to the low-energy limit.
For example, from a Planck-scale perspective, our laboratory
experiments (even the ones conducted at, {\it e.g.} CERN, DESY, SLAC...)
are both low-boost (with respect to the center of mass frame)
and low-energy.
However, the ``UHE cosmic-ray paradox'', for which
a quantum-gravity origin has been conjectured (see later),
occurs in a situation where all the energies of the particles
are still tiny with respect to the Planck energy scale,
but the boost with respect to the center-of-mass frame
(as measured by the ratio $E/m_{proton}$ between the proton energy and
the proton mass) could be considered to be ``large''
from a Planck-scale perspective ($E/m_{proton} \gg E_p/E$).

These concerns are strengthened by looking at the literature
available on  the quantum pictures of spacetime that provide
motivation for the study of modified dispersion relations,
which usually involve
either noncommutative geometry or Loop Quantum Gravity,
where, as mentioned, the outlook of a
low-energy effective-field-theory description is not reassuring.

Of course, in phenomenology this type of concerns can be set aside, since
one is primarily looking for confrontation with experimental data,
rather than theoretical prejudice. It is clearly legitimate to set up
a test theory exploring the possibility of Planck-scale departures
from Lorentz symmetry within the formalism of low-energy effective
field theory. But one must then keep in mind that the implications
for most quantum-gravity research lines of the experimental bounds
obtained in this way might be very limited.
This will indeed be the case if we discover that, as some mentioned
preliminary results suggest,
the limit in which the full quantum-gravity theory reproduces a
description in terms of effective field theory in classical spacetime
is also the limit in which the departures from Lorentz symmetry
must be neglected.

Having provided this long warning, let me now proceed to
a characterization of the test theory which I see emerging from
the works reported in Refs.~\cite{rob,gampul}.
These studies explore the possibility of a linear-in-$L_p$
modification of the dispersion relation
\begin{equation}
 m^2 \simeq E^2 - \vec{p}^2
+  \eta \vec{p}^2 L_p E
~,
\label{dispROB}
\end{equation}
 {\it i.e.} the case $n=1$ of Eq.~(\ref{displeadbis}).
The key assumption in Refs.~\cite{rob,gampul} is that such
modifications of the dispersion relation should be introduced
consistently with an effective low-energy field-theory description
of dynamics.
The implications of this assumption were explored in particular
for fermions and photons\footnote{Actually the studies
in Ref.~\cite{rob,gampul} focus primarily on electrons and photons.
I will assume that the results for photons generalize to all
fermions, but I must warn the reader that the theory work needed
to fully justify this (however natural)
generalization is still in progress.}.
It became quickly clear that in such a setup universality cannot
be assumed, since one must at least accommodate a polarization
dependence for photons: in the field-theoretic setup it turns
out that when right-circular polarized photons satisfy the
dispersion relation $E^2 \simeq p^2 + \eta_\gamma p^3$ then necessarily
left-circular polarized photons satisfy the ``opposite sign"
dispersion relation $E^2 \simeq p^2 - \eta_\gamma p^3$.
For spin-$1/2$ particles the analysis reported in
Ref.~\cite{rob} does not necessarily suggest a
similar helicity dependence, but of course in a context in
which photons experience such a complete correlation of the
sign of the effect with polarization
it would be awkward to assume that instead for
electrons the effect is completely helicity independent.
One therefore introduces two independent parameters $\eta_+$ and $\eta_-$
to characterize the
modification of the dispersion relation for electrons.

These observations provide the basis for a ``GPMP test
theory"\footnote{Here, for the ``GPMP'' short name, I am guided by
the initials of the authors of Refs.~\cite{rob,gampul}, but again
those authors should not be held ``responsible'' for the entire
structure of this GPMP framework I am describing.
Actually, the studies reported in Refs.~\cite{rob} and \cite{gampul} differ
significantly even among them: in particular, while
Ref.~\cite{rob} assumes locality of the new terms in the Lagrangian
density, Ref.~\cite{gampul} also contemplates the possibility of
nonlocality.}.
However, as in the case of the
minimal AEMNS test theory,
it appears wise to first focus the phenomenology on a reduced
two-parameter version
of the test theory, reflecting some natural physical assumptions.
As usual, once the reduced version of the test theory is falsified
one can contemplate its possible generalizations.

In introducing a reduced GPMP test theory
I believe that a key point of naturalness comes from the
observation that the effective-field-theory setup
imposes for photons a modification of the dispersion relation
which has the same magnitude for both polarizations
but opposite sign: it is then natural to give priority to the
hypothesis that for fermions a similar mechanism would apply, {\it i.e.}
the modification of the dispersion relation should have the same
magnitude for both signs of the helicity, but have a correlation
between the sign of the helicity and the sign of the dispersion-relation
modification.
This would correspond to the natural-looking
assumption that the Planck-scale
effects are such that in a beam composed of randomly selected particles
the average speed in the beam is still governed by ordinary special
relativity (the Planck-scale effects average out summing over
polarization/helicity).

A further ``natural'' reduction\footnote{I must here stress that, while
it is ``natural'' to start the phenomenology from the assumption that
all fermions experience the same Planck-scale effects, there
are some ``natural'' mechanisms that could lead to a different
magnitude of the effect for different types of fermions.
As mentioned, one finds an example of such a possibility in the context
of the analysis of certain approaches to field theory in canonical
noncommutative spacetimes.}
of the parameter space
is achieved by assuming that all fermions
are affected by the same modification
of the dispersion relation.

In the following I refer to this reduced two-parameter
GPMP test theory as the ``minimal GPMP test theory'',
characterized by the following ingredients:

(minGPMP.1) right-circular polarized photons
are governed by the dispersion relation
\begin{equation}
 m^2 \simeq E^2 - \vec{p}^2
+  \eta_\gamma \vec{p}^2 \left({E \over E_{p}}\right)
~,
\label{displeadNONUNIgpmpmin1}
\end{equation}
while left-circular polarized photons
are governed by the dispersion relation
\begin{equation}
 m^2 \simeq E^2 - \vec{p}^2
-  \eta_\gamma \vec{p}^2 \left({E \over E_{p}}\right)
~;
\label{displeadNONUNIgpmpmin2}
\end{equation}

(minGPMP.2) for fermions the dispersion relation
takes the form
\begin{equation}
 m^2 \simeq E^2 - \vec{p}^2
+  \eta_f \vec{p}^2 \left({E \over E_{p}}\right)
~,
\label{displeadNONUNIgpmpmin3}
\end{equation}
in the positive-helicity case, while
for negative-helicity fermions
\begin{equation}
 m^2 \simeq E^2 - \vec{p}^2
-  \eta_f \vec{p}^2 \left({E \over E_{p}}\right)
~,
\label{displeadNONUNIgpmpmin4}
\end{equation}
with the same value of $\eta_f$ for all fermions;

(minGPMP.3) dynamics is described in terms of effective low-energy
field theory.

While this ``minimal'' version of the test theory may
deserve to be
the primary focus of GPMP-based phenomenology work, it is of course legitimate
to consider its generalizations with independent parameters
(rather than a single parameter, with the opposite-sign correlation)
for the two helicities of fermions,
and possibly allowing for
different values of the parameters for different species
of fermions.

\section{Some key issues for phenomenology with the two test theories}
I have argued that it is necessary to enter a new more mature phase
of Quantum Gravity Phenomenology, in which the use of some reference
test theories takes center stage.
But I have also observed that there are a number of delicate issues
that need to be considered in setting up such test theories.
In the end, as a way to handle the large variety of scenarios
we should be prepared to face as the understanding of quantum-gravity
pictures progresses, it appeared wise
to introduce both a pure-kinematics test theory
and a field-theory-based test theory.
But, while conceptually these two types of test theories are well motivated
and well defined, one must face additional difficulties in developing a
phenomenology based on these test theories.
For the pure-kinematics test theory the difficulties originate
primarily from the fact that sometimes an effect due to modification
of dynamics can take a form that is not easily
distinguished from a pure-kinematics effect.
For the field-theory-based test theory the difficulties originate
from the fact that the relevant field theory is not renormalizable.

In this section I want to stress these difficulties, but I also
intend to argue that there is a reasonable
way to proceed in spite of these difficulties.

\subsection{On the field-theory-based phenomenology}
In introducing the ``minimal GPMP test theory'' in the previous
section I stressed that
the assumption of a description of dynamics in terms of effective
field theory might fail to capture the insight gained from preliminary
analysis of certain quantum-spacetime pictures.
This ``conceptual concern'' can be easily set aside in phenomenology
work.
However, there is another, potentially more troublesome, issue
that affects phenomenology work with the GPMP test theory:
the relevant field
theory is not renormalizable, and therefore, at least at a strict
in-principle level, it is not predictive.

A description of possible Planck-scale departures from
Lorentz symmetry within effective field theory can only be developed
with a rather strongly pragmatic attitude; in fact,
while one can introduce Planck-scale suppressed effects
at tree level, one
expects that loop corrections would typically lead, for fixed bare parameters
and cutoff scale, to inadmissibly large
departures from ordinary Lorentz symmetry.
The parameters of the theory can be fine-tuned to eliminate
the unwanted large effects, but the needed level of fine tuning
is usually rather unpleasant.
A particularly unpleasant level of fine tuning
might be required in the case of the GPMP test theory
since some authors (notably Refs.~\cite{suda1,suda2}) have argued
that the loop expansion could effectively generate terms
that are unsuppressed by the cut-off scale of the (nonrenormalizable)
field theory.

On the basis of these severe fine-tuning issues one might be tempted
to disregard completely the GPMP test theory.
But I propose that, at a strictly phenomenological level
of analysis, a fine-tuning
problem, however severe, cannot provide sufficient motivation for
disregarding a scenario.
Actually some of the most successful theories used in fundamental
physics are affected by severe fine tuning. Eventually we learn
that the fine tuning is only apparent, that some hidden symmetry was
actually ``naturally'' setting up the hierarchy of parameters.
And it appears that some symmetry principles could also stabilize
the GPMP field theory~\cite{pospeSYMM}.

So I advocate a viewpoint such that the fine-tuning issues do not cause
much concern. But a severe challenge remains: how do we analyze dynamics
with such a nonrenormalizable field theory, affected by troublesome UV
pathologies? Even if we limit the analysis of the GPMP
test theory to tree level, following a strategy which has proven
fruitful in other effective-field-theory approaches,
one could still wonder whether the tree-level analysis should
be limited to dimension-5 operators (as assumed in my description of the
GPMP test theory) or one should also include the other types
of  operators that would be generated through loop effects by those
same dimension-5 terms. I propose that, at the present stage
of phenomenology work, it is legitimate to focus
on the dimension-5 operators. This provides a scenario which can be tested
experimentally, and, as I emphasize later in this paper, experiments
can test this scenario in some detail in the coming years.
So rather than dwell on the specific type of UV sector that would be needed
to stabilize the scenario with dimension-5 operators, we can focus
our efforts on the relevant phenomenology.
If the scenario turns out to be
excluded by data the technical issues become irrelevant,
and if instead the scenario actually turns out to accommodate nicely
some data all the conceptual concerns will be immediately disregarded.

\subsection{On the pure-kinematics phenomenology}
I introduced the minimal AEMNS test theory as the natural starting
point for a phenomenology that prudently focuses on kinematics,
since dynamics is so poorly understood in the relevant quantum-spacetime
pictures.
While this prudent approach may be attractive from a conceptual perspective,
in setting up a phenomenology the idea of focusing on pure-kinematics
tests is very challenging.
The number of contexts in which dynamics does not have an explicit
role is of course very limited, and, even when not immediately evident,
a role for dynamics may easily be hidden somewhere deep in the
analysis.

I propose to handle this challenge by adopting a rather narrow
definition of what a pure-kinematics test should be.
There are clearly at least two aspects of particle-physics analyses
which are truly a reflection of pure kinematics:

\noindent
$\bullet$ the structure of the energy-momentum dispersion relation
(and the associated relation between speed and energy of the particles)
is fixed by pure kinematics, by symmetry principles. By adding new
interactions for a given field on may achieve a shift of the
mass, but the structure of the dispersion relation is protected
by symmetry.

\noindent
$\bullet$ and symmetries also fix the threshold requirements for
particle-physics reaction processes:
no matter which type of interactions are introduced,
if special-relativistic kinematics is assumed
a center-of-mass
collision between two photons can produce and electron positron pair
only if each photon carries energy larger than the mass of electron.

We can therefore perform a pure-kinematics test if we focus
on the analysis of the speed of propagation of particles
or we focus on the energy-threshold requirements for
particle-physics reaction processes.

Concerning the energy-threshold tests which I will discuss later
on a difficulty arises from the fact that sometimes one is not certain
about the energies of the incoming particles, and, as I shall stress,
in some cases an incorrect identification of the energy
of one of the particles could be compensated by a correspondingly
incorrect description of some relevant cross sections.
In this specific sense energy-threshold tests may sometimes involve
a mix between pure-kinematics and dynamics aspects of the theory.
I will attempt to examine this issue in detail in the following.
But, at least from a conceptual perspective,
it is important to realize that genuine
pure-kinematics tests
using energy-threshold studies
can be performed when all the incoming energies
are known.
For example, if in some collisions involving
two photons both with $0.2 MeV$ ($< m_{electron}$)
an electron-positron pair was produced, then special-relativistic
kinematics would be ruled out.

\section{Examples of phenomenological analyses
with the two test theories}
The realization that Planck-scale modifications of the dispersion
relation could lead to observably large effects~\cite{grbgac}
has generated a large research effort over these past few years.
As mentioned most of this work relied on rather preliminary analyses.
The key objective was to demonstrate that indeed Planck-scale
sensitivity could be achieved.
But now that this issue concerning sensitivity is settled,
it is necessary to adopt a more robust style of phenomenology
work. In particular, a meaningful comparison of the sensitivities
achievable with different types of data must rely on the commonly-adopted
language of some reference test theories.
It is of course meaningless to compare limits obtained
within different test theories. And there is no scientific content
in an experimental limit claimed on a vaguely defined test theory.
In the recent literature there has been a proliferation of
papers claiming to improve limits on Planck-scale modifications
of the dispersion relation, but the different studies were simply
considering the same type of dispersion relation within
significantly different (and sometimes not fully characterized)
test theories.

For the reasons discussed in the previous section,
the minimal AEMNS test theory and the minimal GPMP test theory
may provide a good choice of reference test theories.
And their analysis allows to illustrate
that the outlook of a certain class of data
when examined at the level of a test theory may be very different
from what one might expect on the basis of a simplistic sensitivity
estimate.

Besides stressing this point concerning test theories,
I also want to stress that it is important
to be absolutely conservative in assessing the robustness of
the data we are confronted with.
The fact that most of the data relevant for this quantum-gravity
phenomenology comes from astrophysics, and we are therefore not in
the comforting situation of repeated controlled experiments,
should be a forceful motivator for conservative data analyses.

In this section I will consider certain types of
phenomenological analyses that illustrate my proposed strategy
of analysis for the two test theories.

\subsection{Derivation of limits from time-of-flight analyses}\label{grbs}
The best known strategy for establishing experimental limits
on Planck-scale modifications of the dispersion relation
is based~\cite{grbgac}
on the fact that both in
the AEMNS test theory and in the GPMP test theory
one expects a wavelength dependence of the speed of photons,
by combining the modified dispersion relation and
the relation $v = dE/dp$. At ``intermediate energies" ($m < E \ll E_p$)
this velocity law will take the form
\begin{equation}
v \simeq 1 - \frac{m^2}{2 E^2} +  \eta \frac{n+1}{2} \frac{E^n}{E_p^n}
~.
\label{velLIVbis}
\end{equation}
Whereas in ordinary special relativity two photons ($m=0$)
emitted simultaneously would always reach simultaneously a far-away detector,
according to (\ref{velLIVbis}) two simultaneously-emitted
photons should reach the detector at different times
if they carry different energy. Moreover, in the case of the GPMP test
theory even photons with the same energy would arrive at different
times if they carry different polarization.
In fact, while the minimal AEMNS test theory assumes universality,
and therefore
a formula of this type would apply to photons of any polarization,
in the GPMP test theory, as mentioned, the sign of the effect
is correlated with polarization.
As a result, while the AEMNS test theory is best tested by
comparing the arrival times of particles of different energies,
the GPMP test theory is best tested by considering the highest-energy
photons available in the data and looking for a sizeable spread
in times of arrivals (which one would then attribute to
the different speeds of the two polarizations).

This time-of-arrival-difference effect
can be significant~\cite{grbgac,billetal}
in the analysis of short-duration bursts of photons that reach
us from far away sources.

In the near future an opportunity to test this
effect will be provided by observations of gamma-ray bursters.
For a gamma-ray burst it is not uncommon that the time travelled
before reaching our Earth detectors be of order $T \sim 10^{17} s$.
And microbursts within a burst can have very short duration,
as short as $10^{-3} s$ (or even $10^{-4} s$), and this
means that the photons
that compose such a microburst are all emitted at the same time,
up to an uncertainty of $10^{-3} s$.
Some of the photons in these bursts
have energies that extend at least up to the $GeV$ range.
For two photons with energy difference of order $\Delta E \sim 1 GeV$
a $\eta \Delta E/E_p$ speed difference over a time of travel
of $10^{17} s$
would lead to a difference in times of arrival of
order
\begin{equation}
\Delta t \sim \eta T \Delta \frac{E}{E_p} \sim 10^{-2} s
~,
\label{tdelay}
\end{equation}
which is significant (the time-of-arrival differences would be larger
than the time-of-emission differences within a single microburst).

For the AEMNS test theory the
Planck-scale-induced time-of-arrival difference
could be revealed\cite{grbgac,billetal}
upon comparison of the ``average arrival time''
 of the gamma-ray-burst signal (or better a microburst within the
 burst)
in different energy channels.
The GPMP test theory would be most effectively tested by looking
for a dependence of the time-spread of the bursts that grows
with energy (at low energies the effect is anyway small, so the
polarization dependence is ineffective, whereas at high energies
the effect may be nonnegligible and an overall time-spread of
the burst could result from the dependence of speed on
polarization).

Since the quality of relevant gamma-ray-burst data
is still relatively poor,
the present best limit was obtained
in Ref.~\cite{billetal}: the negative results
of a search of time-of-arrival/energy correlations
for a TeV-gamma-ray short-duration flare from the
Markarian 421 blazar
allowed to deduce the limit $|\eta| < 3 {\cdot} 10^{2}$.
Assuming that the relevant gamma-ray emission was not
largely polarized, one would  correspondingly obtain for
the minimal GPMP test theory $|\eta_\gamma| < 1.5 {\cdot} 10^{2}$
(the factor-2 difference in sensitivities for $|\eta|$
and for $|\eta_\gamma|$ is due to the fact that there is
a $2|\eta_\gamma|$ speed-difference effect between polarizations).
However, to my knowledge, the possibility of
large polarization of the
relevant gamma-ray emission (while unexpected)
is not excluded by data, and therefore for the minimal
GPMP test theory these data do not allow to establish
a fully robust limit.

The sensitivities achievable~\cite{glast}
with the next generation of gamma-ray telescopes,
such as GLAST~\cite{glast},
could allow to test very significantly (\ref{velLIVbis})
in the case $n=1$, by possibly pushing the
limit on $\eta$ far below $1$ (whereas the effects found in the
case $n=2$, $|\eta| \sim 1$ are too small for GLAST).
Whether or not these levels of sensitivity
to the Planck-scale effects are actually achieved
may depend on progress in understanding other aspects of
gamma-ray-burst physics.
In fact, the Planck-scale-effect
analysis would be severely affected if there
were poorly understood at-the-source correlations between
energy of the photons and time of emission.
In the recent Ref.~\cite{piranKARP} it was emphasized that
it appears that one can infer such an energy/time-of-emission
correlation from available gamma-ray-burst data.
The studies of Planck-scale effects will be therefore confronted
with a severe challenge of ``background/noise removal''.
At present it is difficult to guess whether this problem
can be handled successfully. We do have a good card to play
in this analysis: the Planck-scale picture predicts that
the times of arrival should depend on energy in a way that
grows in exactly linear way with the distance of the source.
One may therefore hope that, once a large enough sample
of gamma-ray bursts (with known source distances) becomes available,
one might be able
disentangle the Planck-scale propagation effect
from the at-the-source background.

An even higher sensitivity to possible Planck-scale
modifications of the velocity law could be achieved
by exploiting the fact that, according to
current models~\cite{grbNEUTRINOnew},
gamma-ray bursters should also emit a substantial amount of
high-energy neutrinos.
Some neutrino observatories should soon observe neutrinos with energies
between $10^{14}$ and $10^{19}$ $eV$, and one could, for example, compare
the times of arrival of these neutrinos emitted by
gamma-ray bursters to the corresponding times of arrival of
low-energy photons.
One could use this strategy to test rather stringently\footnote{Note
however that in an analysis mixing the properties of different particles
the sensitivity that can be achieved will depend strongly on whether
universality of the modification of the dispersion relation
is assumed.
For example, for the GPMP test theory
a comparison of times of arrival of neutrinos and photons
could only introduce a bound on some combination of the
dispersion-relation-modification parameters for the photon and for the
neutrino sectors.}
the case
of (\ref{velLIVbis}) with $n=1$, an even perhaps gain some access to
the investigation of the case $n=2$.

In order to achieve these sensitivities with neutrino studies
once again
some technical and conceptual challenges should be
overcome. Also this type of analysis requires
an understanding
of gamma-ray bursters good enough to establish whether there are typical
at-the-source time delays. The analysis would loose much of its potential
if one cannot exclude some systematic tendency of
gamma-ray bursters to emit high-energy
neutrinos with, say, a certain delay with
respect to microbursts of photons. But also in this case
one could hope to combine several observations
from gamma-ray bursters at different distances in order
to disentangle the possible at-the-source effect.

\subsection{Analysis of threshold-energy requirements
in the laboratory}
As mentioned, in addition to
the possible manifestation in time-of-arrival/energy correlations,
the quantum-gravity-scale modifications of the dispersion relation
could have
observably-large implications for what concerns the
analysis of energy threshold for particle-physics reactions.
This possibility has been studied primarily in contexts of
interest in astrophysics~\cite{kifu,ita,aus,gactp},
where the relevant scales turn out to be favourable for
achieving high sensitivity.
But before commenting on those analyses in astrophysics,
I find useful to consider the study of these processes
from a gedanken-experiment perspective, imagining to perform
analogous studies in the controlled environment of
a laboratory setup.

I should first of all emphasize that I will focus on the possibility
that
the modification of the dispersion relation occurs in a framework
in which the law of energy-momentum conservation is not modified.
Both the AEMNS test theory and the GPMP test theory involve
modified dispersion relations and unmodified laws of energy-momentum
conservation (the fact that the law of energy-momentum conservation
is not modified is explicitly among the ingredients of the AEMNS
test theory, while in the GPMP test theory it follows from
the adoption of low-energy effective field theory).

In this paper I am not discussing in detail the case of
modified dispersion relations introduced within a ``doubly-special
relativity'' scenario~\cite{gacdsr,dsrmost1,dsrmost2},
which I already mentioned in the discussion of $\kappa$-Minkowski
spacetime.
I am in fact here focusing on scenarios for broken Lorentz
symmetry (rather than deformed Lorentz symmetry).
Test theories for doubly-special
relativity scenarios with modified dispersion relations
are under consideration (see, {\it e.g.}, Ref.~\cite{sethdsr,dsrphen}),
but I will not make room for them here. It is appropriate however
to stress in this subsection that the assumption of
modified dispersion relations and unmodified laws of energy-momentum
conservation is inconsistent with the doubly-special
relativity principles, since it inevitably~\cite{gacdsr} gives rise to
a preferred class of inertial observers.
A doubly-special
relativity scenario with modified dispersion relations must necessarily
have a modified law of energy-momentum conservation.

Going back to the AEMNS and GPMP test theories which I am considering,
in this subsection I want to stress that
combining a modified dispersion relation
with unmodified laws of
energy-momentum conservation
one naturally finds a modification
of the threshold requirements for
certain reactions.
Let us in particular consider the dispersion relation (\ref{displeadbis}),
with $n=1$, for the AEMNS test theory, in the analysis of a
process $\gamma \gamma \rightarrow e^+ e^-$,
a collision between
a soft photon of energy $\epsilon$
and a high-energy photon of energy $E$ which might produce an
electron-positron pair.
For given soft-photon energy $\epsilon$,
the process $\gamma \gamma \rightarrow e^+ e^-$
is allowed only if $E$ is greater than a certain
threshold energy $E_{th}$ which depends on $\epsilon$ and $m_e^2$.
For $n=1$, combining (\ref{displeadbis}) with unmodified
energy-momentum conservation,
this threshold energy
(assuming $\epsilon \ll m_e \ll E_{th} \ll E_p$)
must satisfy
\begin{equation}
E_{th} \epsilon + \eta \frac{E_{th}^3}{8 E_p} \simeq m_e^2
~.
\label{thrTRE}
\end{equation}
The special-relativistic result $E_{th} = m_e^2 /\epsilon$
corresponds of course to the $\eta \rightarrow 0$ limit
of (\ref{thrTRE}).
For $|\eta | \sim 1$ the Planck-scale correction can be
safely neglected as long as $\epsilon > (m_e^4/E_p)^{1/3}$.
But eventually, for sufficiently small values of $\epsilon$
(and correspondingly large values of $E_{th}$) the
Planck-scale correction cannot be ignored.

This is another pure-kinematics test: if a $10 TeV$ photon
collides with a photon of $0.03 eV$ and produces
an electron-positron pair the case $n=1$, $\eta \sim -1$
for the AEMNS is ruled out.
A $10 TeV$ photon and a $0.03 eV$ photon can produce an
electron-positron pair according to ordinary special-relativistic
kinematics (and its associated requirement $E_{th} = m_e^2 /\epsilon$),
but they cannot produce an
electron-positron pair according to AEMNS kinematics (and its associated
requirement (\ref{thrTRE})).

So for negative $\eta$ the AEMNS test theory could be ruled relying
on pure-kinematics data.
For positive $\eta$ the situation is somewhat different.
While negative $\eta$ increases the energy requirement for electron-positron
pair production, positive $\eta$ decreases
the energy requirement for electron-positron
pair production.
In some cases where one would expect
electron-positron
pair production to be forbidden the AEMNS test theory with positive $\eta$
would instead allow it. But once a process is allowed there is no
guarantee that it will actually occur, not without some information
on the description of dynamics (that allows us to evaluate cross sections).
So, from a fully conservative perspective, a pure-kinematics framework
can be falsified when it predicts that a process cannot occur
(if instead the process is seen) but it cannot be falsified when
it predicts that a process is allowed.

In the case of the minimal GPMP test theory the availability of
the field-theoretic setup for the description of dynamics
renders this types of studies even more powerful
(when the GPMP predicts that a process is allowed it also predicts
the relevant probability amplitudes).
Also for tests of the minimal GPMP test theory
in the laboratory one could use
a beam of $10 TeV$ photon and a beam of $0.03 eV$ photons.
However, for the GPMP test theory it would be very useful
to control helicity/polarization of the beams.

It is actually not so inconceivable~\cite{gactp}
to conduct this type of tests really
in a laboratory (since the production of 10-TeV photons
is not so far from our present technical capabilities),
but this might have to wait a few years.
In the meantime some contexts in astrophysics
provide us an opportunity to study the relevant processes,
although rather indirectly and without the comfort level of
a controlled laboratory setup.

\subsection{Limits obtained from observed absorption
of TeV photons from Blazars}
The type of threshold-energy scenario discussed
in the preceding subsection
could have~\cite{kifu,ita,aus,gactp}
observably-large implications for what concerns the opacity
of our Universe to various types of high-energy particles.
Of particular interest
is the fact that, according to the conventional (classical-spacetime)
description,
the infrared diffuse extragalactic background
should give rise to strong absorption of ``$TeV$ photons"
(here understood as photons with energy $1 TeV < E < 30 TeV$).
The relevant process is of course $\gamma \gamma \rightarrow e^+ e^-$,
already analyzed in the preceding subsection.

If the photon of energy $\epsilon$ is part of the
infrared diffuse extragalactic background
and the photon  emitted by
a blazar is of TeV-range energy
one finds that the prediction for
absorption of the hard photon by the
infrared diffuse extragalactic background
can be significantly modified.

The classical-spacetime analysis, in which a key role is played
by the threshold condition $\epsilon \geq m_e^2/E$,
the distance of the blazar,
and the density of
the infrared diffuse extragalactic background,
leads to a prediction
of the amount of absorption to be expected as a function
of the energy of the photons emitted by the blazar.
The experimental verification of this classical-spacetime prediction
has made significant progress over the last couple of years:
evidence of absorption of $TeV$ photons
has been reported in observations~\cite{krennMKcut,ahaMKcut}
of the Markarian 421 blazar (at a redshift of $z=0.031$),
in observations~\cite{ahaMKcut501} of
the Markarian 501 blazar (at a redshift of $z=0.034$),
and in observations~\cite{h1426} of the blazar H1426+428
(at a redshift of $z=0.129$).
While these observations all concerned $\gamma$-rays up to
energies in the 20-$TeV$ range,
observations of $\gamma$-rays up to
45 TeV from Markarian 421 have been recently reported~\cite{tev45},
and again the data are found~\cite{tev45} to reflect significant
absorption.

Perhaps the most convincing evidence of
TeV-photon absorption come from
the analysis of the combined X-ray/TeV-gamma-ray spectrum
for the Markarian 421 blazar, as discussed in particular in
Ref.~\cite{steckXRAY}.
The X-ray part of the spectrum allows to predict the
TeV-gamma-ray part of the spectrum
in a way that is rather insensitive on our
poor knowledge of the source. This in turn allows
to establish in a source-independent way that absorption
is occurring.

The fact that
the observations still give us only a preliminary picture of absorption
together with the fact that there is a significant level of
uncertainty in phenomenological models of $TeV$ blazars
and in phenomenological models of the
density of the infrared diffuse extragalactic background
does not allow us to convert these observations into tight limits
on departures from the classical-spacetime analysis.
However, I intend to argue
that even just the basic fact
that we see absorption of $TeV$ $\gamma$-rays allows to
derive a rather robust limit.

Previous studies (see, {\it e.g.}, Refs.~\cite{ita,glasteck,tedtwo})
had already shown that this type of observations, if found to
be in agreement with the conventional classical-spacetime picture,
could constrain very significantly several models of
departure from Lorentz symmetry.
In line with these previous studies, and using the fact
that the observations recently reported in Ref.~\cite{tev45}
further extend the energy range of observations
of $TeV$ blazars, one can easily verify that
Planck-scale sensitivity (intended as $\eta \sim 1$ sensitivity,
at least for $n=1$) is within reach for the near future.
However, my main focus is not on this type of future-sensitivity estimate,
but rather on the limit
that can be conservatively/robustly established using
presently-available information, {\it i.e.} using the bare
fact that some absorption of $TeV$ $\gamma$-rays is evident in the data.

In fact, while the presence of some level of absorption
of $TeV$ $\gamma$-rays is indeed evident in the observations
reported in Refs.~\cite{krennMKcut,ahaMKcut,ahaMKcut501,h1426,tev45},
these observations are still insufficient to make a quantitative
comparison with the predictions of the classical-spacetime picture,
at least not consistently with the prudently conservative attitude one must
adopt in attempting to
establish an unconditional limit on parameters such as $\eta$.
The fact that some absorption
of $TeV$ $\gamma$-rays is being seen can be robustly inferred
from the structure of all of the observations
reported in Refs.~\cite{krennMKcut,ahaMKcut,ahaMKcut501,h1426,tev45}.
On the other hand if one looks in detail at the information that
is emerging from these observations it is rather clear that
we are not ready for stating robustly that the predictions
of the classical-spacetime picture are finding
detailed confirmation:

\begin{list}{}{}

\smallskip

\item (j) Some authors have
discussed~\cite{krennMKcut,ahaMKcut,ahaMKcut501,voelk}
a puzzling difference between the cutoff energy found
in data concerning Markarian 421, $E_{mk421}^{cutoff} \simeq 3.6 TeV$,
and the corresponding cutoff estimate obtained
from Markarian 501, $E_{mk501}^{cutoff} \simeq 6.2 TeV$,
a difference which appears to be significant at the $3 \sigma$
level.
Since Markarian 421 and  Markarian 501 are at
comparable distances from the Earth (at redshifts of $z=0.031$
and $z=0.034$ respectively)
and they are
expected to host very similar mechanisms of
emission of $TeV$ $\gamma$-rays,
this difference in the estimated cutoff energy
may be an indication that we do not yet have a robust picture
of what is going on.

\smallskip

\item (jj) The observation of $TeV$ $\gamma$-rays emitted by
the blazar H1426+428, which is at a redshift four times bigger
than the one of Markarian 421 and  Markarian 501, does show,
as expected in the standard picture, a level of absorption
which is higher than the ones inferred for Markarian 421
and  Markarian 501. However, as emphasized in Ref.~\cite{h1426},
even taking into account the uncertainties on the density of
the infrared diffuse extragalactic background,
 ``the TeV luminosity seems to
exceed the level anticipated from the current models of TeV blazars
by far"~\cite{h1426}.

\smallskip

\item (jjj) As mentioned, a detailed comparison of observed absorption
with corresponding predictions
of the classical-spacetime (Lorentz-invariant)
description of absorption
by the infrared diffuse extragalactic background
of $\gamma$-rays emitted by blazars would require
correspondingly accurate descriptions of the spectrum emitted
by the blazars and of the density
of the infrared diffuse extragalactic background.
However, measurements of the
density of the infrared diffuse extragalactic background
are very difficult
and as a result our experimental information
on this density is still affected by
large uncertainties~\cite{voelk,berezin}.
Similarly, there are models of $TeV$ blazars which appear to
be rather robust theoretically, but some of the above-mentioned
observational facts (the different cutoff estimates for Markarian 421
and Markarian 501 and the unexpectedly large $TeV$ luminosity
of the H1426+428 blazar) impose us to treat cautiously the
indications obtained from these theoretical models.

\end{list}

These points (j), (jj), (jjj) impose us to analyze prudently
the implications of the observations reported
in Refs.~\cite{krennMKcut,ahaMKcut,ahaMKcut501,h1426,tev45}.
I shall not assume that the observations imply any level
of agreement with the classical spacetime picture,
but I will insist that the Planck-scale effect
be consistent with the fact, now established,
that $TeV$ $\gamma$-rays with energies up to 20 $TeV$
are absorbed by the infrared diffuse extragalactic background.
This suggests that at least some photons with energy
smaller than $\sim 200 meV$
can create an electron-positron pair
in collisions with a $20 TeV$ $\gamma$-ray.
For the AEMNS test theory, in light of Eq.~(\ref{thrTRE}),
this observation leads to
\begin{equation}
\eta \geq - 46
~
\label{newlimitCONSERV}
\end{equation}
({\it i.e.} either $\eta$ is positive or $\eta$ is negative with
absolute value smaller than 46).

Other authors~\cite{glasteck,steckEXTRA,tedtwo}
have argued that a certain level of agreement with the predictions
of the classical-spacetime picture can be inferred from the data,
and on that basis they have derived more stringent limits
than the one I am claiming in (\ref{newlimitCONSERV}).
However, I am here insisting on a concept of experimental limit
that is absolutely conservative, an experimental limit that
can be truly considered as an unavoidable fact to be taken into
account by theorists. For the reasons discussed above any claim
that there is some agreement in the observed absorption and
the level of absorption predicted by the classical-spacetime
picture would be {\underline{conditional}} to the success of
some, still unproven, models of TeV-gamma-ray emission by blazars
and of the infrared diffuse extragalactic background.

Actually, one might argue that
even the more prudent limit (\ref{newlimitCONSERV}) I am advocating
should be subject to further scrutiny.
In fact, we see absorption of the multi-TeV gamma rays
and we assume it should be due to interactions with infrared photons;
however, one could conjecture that perhaps
the absorption is due to higher-energy background photons.
Since we intend to confine the analysis of the AEMNS test theory
to the realm of pure kinematics, any bound that is established
on the parameters of the AEMNS test theory should be completely
insensitive on dynamics issues.
We should therefore contemplate the possibility that
the AEMNS kinematics be implemented within a framework
in which the description of dynamics is such to introduce
a large-enough modification of cross sections to allow absorption
of multi-TeV blazar gamma rays by background photons of energy
higher than  $200 meV$.
This would be a way to evade the bound (\ref{newlimitCONSERV}).
Consistently with the absolutely conservative
approach I am advocating I will therefore not describe
(\ref{newlimitCONSERV}) as a fully-established experimental bound.
I propose however that (\ref{newlimitCONSERV}) is rather robust:
the only way to evade it requires a sort of careful ``conspiracy''
of new effects.
In the standard picture photons with below-TeV energy should not be
absorbed while multi-TeV photons should be absorbed by infrared photons.
This picture is of course also consistent with AEMNS kinematics,
if $\eta$ satisfies the requirement (\ref{newlimitCONSERV}).
In order to allow negative values of $\eta$ of absolute value larger
than prescribed by (\ref{newlimitCONSERV}) one would need a
corresponding modification of cross sections, so that the modification
of kinematics and the modification of cross sections would conspire
to leave the numerical value of the scale of onset of absorption
basically unchanged with respect to the standard picture.
It is hard to believe that such a conspiracy would be in place,
but I will still describe the limit (\ref{newlimitCONSERV})
as established ``up to conspiracies''.

So far in this subsection my
derivation and discussion of the
limit (\ref{newlimitCONSERV})
is strictly applicable only to the minimal AEMNS test theory.
For the case of the minimal GPMP test theory
the analysis is simplified by the fact that one has the field-theoretic
setup for the evaluation of cross sections,
but on the other hand one has the complication of having
to take into account the fact
that the modification of the dispersion relation carries
opposite sign for the two polarizations of the photon
and for the two helicities of the electron/positron.
And one should also take into account that,
while observations now provide
robust evidence of some absorption of TeV gamma rays,
a conservative phenomenological analysis should
(as a result of the mentioned residual grey areas
in the understanding of these observations)
consider the possibility that only one of the polarizations
is being absorbed.
I postpone this more involved analysis to a future study.

\subsection{Derivation of limits from analysis of UHE cosmic rays}\label{uhecr}
In the preceding subsection I discussed the implications of possible
Planck-scale effects
for the process $\gamma \gamma \rightarrow e^+ e^-$,
but of course this is not the only process in which Planck-scale
effects can be important. In particular, there has been strong
interest\cite{kifu,ita,aus,gactp,tedtwo,gacpion,orfeupion,nguhecr}
in ``photopion production", $p \gamma \rightarrow p \pi$,
where again the combination of
(\ref{displeadbis}) with unmodified
energy-momentum conservation
leads to a modification of the minimum proton energy required
by the process (for given photon energy).
In the case in which the photon energy is the one typical of CMBR photons,
in the AEMNS test theory
one finds that the threshold proton energy can be significantly shifted
upward (for negative $\eta$), and this
in turn should affect at an observably large level the
expected ``GZK cutoff" for the observed cosmic-ray spectrum.
Observations reported by the AGASA\cite{agasa} cosmic-ray
observatory provide some encouragement for the idea of
such an upward shift of the GZK cutoff, but the issue
must be further explored.
Forthcoming cosmic-ray observatories, such as Auger\cite{auger},
should be able\cite{kifu,gactp} to fully investigate this possibility.

Of course, also for the cosmic-ray GZK threshold, just like for the
gamma-ray absorption threshold discussed in the preceding subsection,
the AEMNS analysis should contemplate the possibility of
a ``conspiracy'', although in this case it appears to be an ubelievable
conspiracy.
If the only background radiation available for photopion production
was the CMBR, then
the prediction of an upward shift of the GZK
cosmic-ray cutoff within the AEMNS test theory,
for negative $\eta$, would be completely robust.
But background radiation has many components and one could again
(as in the case of the
gamma-ray absorption threshold) contemplate the possibility
to combine AEMNS kinematics with an unspecified
description of dynamics such
that interactions of cosmic rays with other components of the
background radiation would lead to a net result that does not change
the numerical value of the GZK threshold.
At least for $n=1$ and negative $\eta$ of order 1,
I advocate that
this ``conspiracy scenario'' should be dismissed.
For $n=1$ and negative $\eta$ of order 1
the AEMNS kinematics allows the interaction of cosmic rays
only with photons of  energy higher than the TeV scale
(see Ref.~\cite{gactp}).
The density of such high energy background photons is
extremely low, and therefore, even in a prudent phenomenology,
this ``conspiracy scenario'' can indeed be dismissed.

For the minimal
GPMP test theory this issue of possible conspiracies is of course
absent, since the field-theoretic setup allows to evaluate cross sections,
but one must take into account that
for one of the helicities of the proton the dispersion relation
is of negative-$\eta$ type while for the other
helicity the dispersion relation
is of positive-$\eta$ type.
One would then expect roughly one half of the UHE protons to evade
the GZK cutoff, so the cutoff would still be violated but in
a softer way than in the case of the AEMNS test theory with
negative $\eta$.

\subsection{Derivation of limits from analysis of photon stability}
The cases considered in the preceding two subsections,
the one of TeV-gamma-ray photon absorption and the one of
photopion production, are examples
of situations in which a given process is allowed
in presence of exact Lorentz symmetry but can be
kinematically forbidden in presence of certain
departures from Lorentz symmetry.
The opposite is also possible: some processes that are
kinematically forbidden in presence of exact
Lorentz symmetry become kinematically allowed
in presence of certain
departures from Lorentz symmetry.

Certain observations in astrophysics,
which allow us to establish
that photons of energies up to $\sim 10^{14}eV$
are not unstable,
can be particularly useful~\cite{tedtwo,gacpion,orfeupion,seth}
in setting limits on some schemes for departures
from Lorentz symmetry.
Let us for example analyze the process $\gamma \rightarrow e^+ e^-$
from the AEMNS perspective,
using the dispersion relation (\ref{displeadbis}), with $n=1$,
and unmodified energy-momentum conservation.
One easily finds a relation between
the energy $E_\gamma$ of the incoming photon, the opening angle $\theta$
between the outgoing electron-positron pair, and the energy $E_+$ of
the outgoing positron (of course the energy of the outgoing electron
is simply given by $E_\gamma - E_+$).
For the region of phase space with $m_e \ll E_\gamma \ll E_p$
this relation takes the form
\begin{eqnarray}
\cos(\theta) &\! \simeq \!& \frac{E_+ (E_\gamma -E_+) + m_e^2
- \eta  E_\gamma E_+ (E_\gamma -E_+)/E_p}{ E_+ (E_\gamma -E_+)} ~,
\label{gammathresh}
\end{eqnarray}
where $m_e$ is the electron mass.

The fact that for $\eta = 0$ Eq.~(\ref{gammathresh}) would
require $cos(\theta) > 1$ reflects the fact that, if Lorentz symmetry
is preserved, the process $\gamma \rightarrow e^+ e^-$ is kinematically
forbidden. For $\eta < 0$ the process is still forbidden, but for
positive $\eta$ high-energy photons can decay
into an electron-positron pair. In fact,
for $E_\gamma \gg (m_e^2 E_p/|\eta |)^{1/3}$
one finds that
there is a region of
phase space where $\cos(\theta) < 1$, {\it i.e.} there is a physical
phase space available for the decay.

The energy scale $(m_e^2 E_p)^{1/3} \sim 10^{13} eV $ is not
too high for testing, since, as mentioned, in
astrophysics we see photons of energies up to $\sim 10^{14}eV$
that are not unstable (they clearly travel safely some large astrophysical
distances).

Within AEMNS kinematics, for $n=1$ and
positive $\eta$ of order 1, it would have been natural
to expect that such photons with $\sim 10^{14}eV$ energy
would not be stable.
Once again, before claiming that $n=1$ and
positive $\eta$ of order 1 is ruled out, one should be concerned
about possible conspiracies.
The fact that the decay of $10^{14}eV$ photons is allowed
by AEMNS kinematics (for $n=1$ and
positive $\eta$ of order 1) of course does not guarantee
that these photons should rapidly decay. It depends on the relevant
probability amplitude, whose evaluation goes beyond the reach
of kinematics.
I am unable to provide an intuition
for how big of a conspiracy would be needed
to render $10^{14}eV$ photons stable compatibly with
AEMNS kinematics with $n=1$ and $\eta = 1$.
My tentative conclusion is that $n=1$ with
positive $\eta$ of order 1 is ruled out ``up to conspiracies'',
but unlike the case of the GZK-threshold analysis I am unprepared
to argue that the needed conspiracy is truly unbelievable.

For the GPMP test theory the photon stability analysis
is weakened because of other reasons. There one does have the support
of the effective-field-theory descrition of dynamics, and within that
framework one can exclude huge suppression by Planck scale effects
of the interaction vertex needed for $\gamma \rightarrow e^+ e^-$
around $\sim 10^{13}eV$, $\sim 10^{14}eV$.
So the limit-setting effort is not weakened by the absence
of an interaction vertex. However, as mentioned, consistency with
the effective-field-theory setup
requires that the two polarizations of the photon acquire
opposite-sign modifications of the dispersion relation.
We observe in astrophysics some photons
of energies up to $\sim 10^{14}eV$
that are stable over large distances,
but as far as we know those photons could be all,
say, right-circular polarized (or all left-circular polarized).
I postpone a detailed analysis to future work, but let me note here
that\footnote{I thank an anonimous referee for bringing this
point to my attention.}
there is a region of minimal-GPMP parameter space where both
polarizations of a $\sim 10^{14}eV$
photon are unstable (a subset of the region
with $|\eta_f|>|\eta_\gamma|$). That region of parameter space
is of course excluded by the photon-stability data.

\subsection{Derivation of limits from analysis of synchrotron radiation}
A recent series of
papers\cite{jacoNATv1,newlimit,jaconature,tedreply,carrosync,nycksync,tedsteck}
has focused on the possibility to set limits on Planck-scale modified
dispersion relations focusing on their implications for synchrotron radiation.
By comparing the content of the first estimates\footnote{Ref.~\cite{jacoNATv1}
is at this point obsolete, since the relevant manuscript
has been revised for the published version\cite{jaconature}
and the recent Ref.~\cite{tedsteck} provides an even more
detailed analysis. It is nevertheless useful to consider
this series of manuscripts~\cite{jacoNATv1,jaconature,tedsteck}
as an illustration of how much the outlook of a phenomenological
analysis may change in going from the level of
simplistic order-of-magnitude
estimates to the level of careful comparison with
meaningful test theories.} produced in
this research line~\cite{jacoNATv1}
with the understanding that emerged from follow-up
studies~\cite{newlimit,jaconature,tedreply,carrosync,nycksync,tedsteck}
one can gain valuable insight on the risks involved in analyses based on
simplistic order-of-magnitude
estimates, rather than
careful comparison with
meaningful test theories.
In Ref.\cite{jacoNATv1} the starting point is the observation
that in the conventional (Lorentz-invariant) description of synchrotron
radiation one can estimate the characteristic energy $E_c$ of
the radiation through a heuristic analysis~\cite{jackson}
leading to the formula
\begin{equation}
E_c \simeq {1 \over
R {\cdot} \delta {\cdot} [v_\gamma - v_e]}
~,
\label{omegacjack}
\end{equation}
where $v_e$ is the speed of the electron,
$v_\gamma$ is the speed
of the photon, $\delta$ is the angle of outgoing radiation,
and $R$ is the radius of curvature of
the trajectory of the electron.

Assuming that the only Planck-scale modification in this formula
should come from the velocity law (described using $v=dE/dp$
in terms of the modified dispersion relation),
one finds that in some instances the characteristic energy of
synchrotron
radiation may be significantly modified by the presence of
Planck-scale departures from Lorentz symmetry.
As an opportunity to test such a modification of the
value of the synchrotron-radiation characteristic energy one
can hope to use some relevant data~\cite{jacoNATv1,jaconature}
on photons detected from the Crab nebula.
This must be done with caution since
the observational information on synchrotron radiation being emitted
by the Crab nebula is rather indirect: some of the photons we observe
from the Crab nebula are attributed to sychrotron processes on the basis
of a rather successful model, and the value of the
relevant magnetic fields is also not directly measured.

Assuming that indeed the observational situation has been properly
interpreted, and relying on the mentioned assumption that
the only modification to be taken into account is the
one of the velocity law,
one could basically rule out~\cite{jacoNATv1} the case $n=1$
with negative $\eta$
for a modified dispersion relation
of the type (\ref{displeadbis}).

This observation led at first to some excitement,
but more recent papers are starting to adopt a more prudent viewpoint.
The lack of
comparison with a meaningful test theory represents a
severe limitation of the original analysis.
In particular, synchrotron radiation is due to the acceleration
of the relevant electrons and therefore implicit
in the derivation of the formula (\ref{omegacjack})
is a subtle role for dynamics~\cite{newlimit}.
From a field-theory perspective the process of
synchrotron-radiation emission
can be described in terms
of Compton scattering of the electrons
with the virtual photons of the magnetic field.
One would therefore
be looking deep into the dynamical features of the theory.

The minimal AEMNS test theory does assume a
modified dispersion relation
of the type (\ref{displeadbis}) universally applied
to all particles,
but it is a pure-kinematics framework and, since the analysis
involves some aspects of dynamics,
it cannot be tested using a Crab-nebula
synchrotron-radiation analysis.
I have stressed that also in other instances, like the analysis of the
cosmic-ray GZK threshold and the analysis of the gamma-ray absorption
threshold, there is a possible (conspiracy-type) hidden role of dynamics
in the AEMNS analysis, but in this synchrotron-radiation context
the role of dynamics is explicit and unavoidable.
For example, the concerns about dynamics in the
analysis of the gamma-ray absorption
threshold are only an accident due to the fact that we have no control
over the radiation background.
If one could set up controlled laboratory collisions
between $10 TeV$ photons and $0.03 eV$ photons,
then a pure-kinematics analysis could be truly performed,
without any risk of ``contamination'' from dynamics.
Instead even the study of synchrotron radiation in a controlled
laboratory setup could not be used as a pure-kinematics test:
in the laboratory we can switch on the electron beam and the external
magnetic field, but we have no control on the description/nature
of virtual particles.

For what concerns the minimal GPMP test theory,
where the dynamical aspects of the problem are handled according
to the field-theoretic setup,
the usefulness of this Crab-nebula synchrotron-radiation analysis
is reduced by the fact that we do not know whether both helicities
of the electron (or positron)
are contributing to the synchrotron-radiation
emission.
Through the Crab-nebula synchrotron-radiation analysis
one therefore obtains no constraint on the
minimal GPMP test theory.
The Crab-nebula synchrotron-radiation analysis
can however, as stressed in Ref.~\cite{tedsteck},
introduce a valuable constraint
on more general formulations of the GPMP test theory,
in which one accommodates independent free parameters
for the dispersion-relation modifications of the
two helicities of a fermion.

\section{Some other opportunities to constrain parameters
of the test theories}\label{secn}
As stressed earlier,
the primary objective of this paper is the one of igniting
a transition to a new more mature phase of quantum-gravity
phenomenology, in which different works are compared
in terms of the common language of some reference test theories.
I discussed two ``minimal'' test theories that could be considered
for this purpose, and, in the preceding section, I illustrated
the type of issues that can arise in working with these two test theories
in the context of a few among the most popular
opportunities to test Planck-scale departures from Lorentz symmetry.
While for my purposes it was sufficient to discuss a few examples
of analyses of the test theories, without attempting to provide
a ``status report'' on the absolute best limits achievable
with presently-available data,
in this section I do want to comment briefly on some other
possible opportunities to constrain the two minimal test theories.

Let me start by mentioning that both the AEMNS and GPMP test theories
are preferred-frame theories, and their consistency with the
relevant classic tests ({\it e.g.} Hughes-Drever tests)
should be examined.
Those working in the field have always quickly assumed
that, because of the low energies of the particles involved
in those tests, the corresponding experimental bounds would
be of marginal significance.
This type of low-energy tests usually constrains more effectively
deformations that are not suppressed at low energies ({\it e.g.}
deformations that in field-theory language correspond to dimension-4
operators, rather than the dimension-5 operators of the GPMP test theory).
However,
since the field has now reached a certain maturity, it appears
that a careful analysis of these ``preferred-frame tests''
should be among the priorities for future developments.

For what concerns
specifically the parameter $\eta_f$ of the minimal GPMP test theory,
it has been argued in Ref.~\cite{rob}
that through the results of measurements of spin-polarized
torsion-pendulum frequencies~\cite{etaflim}
one can establish $|\eta_f| \leq 2$.

Concerning particle decays I should mention that, while in some
cases departures from Lorentz symmetry allow the decay
of stable particles (as in the discussed $\gamma \rightarrow e^+ e^-$
context),
it is also possible for
departures from Lorentz symmetry of the type codified in the
minimal AEMNS and minimal GPMP test theories to render
stable, at ultrahigh energies, a particle which would be unstable
in the standard framework.
In particular, there has been some interest~\cite{gacpion,orfeupion}
in the possibility that the process $\pi \rightarrow \gamma \gamma$ might
be forbidden at ultrahigh energies.

Another opportunity that has generated interest recently and I have
not mentioned so far, is the one of
the vacuum Cerenkov
constraint, analyzed in the sense considered for example
in Ref.~\cite{tedsteck}.
This is particularly of interest for some generalizations
of the minimal GPMP
test theory, where the
vacuum-Cerenkov constrain and the synchrotron-radiation
constrain can be considered in an overall analysis~\cite{tedsteck}
of Crab-nebula data.
The effectiveness of this overall analysis may be reduced
by the fact that, as acknowledged in the published version of
Ref.~\cite{tedsteck}, one must consider even the possibility that
the Crab-nebula synchrotron radiation be due to positron acceleration,
but the analysis is valuable nonetheless~\cite{tedsteck}.

For the GPMP test theory\footnote{Clearly the minimal AEMNS test
theory, with its universal modification
of the dispersion relation, predicts no birefringence
effects.} perhaps the best opportunity to constrain the
parameter $\eta_\gamma$ comes from
birefringence analyses:
according to the minimal GPMP test theory
electromagnetic waves of opposite helicity should have different
phase velocities~\cite{gampul,tedsteck,gleiser}.
As the electromagnetic wave travels its linear polarization
should rotate direction as a linear function of time travelled.
Experimental limits on this effect can
be derived using observations of polarized light
from distant galaxies~\cite{tedsteck,gleiser,jackbire}.
The analysis reported in Ref.~\cite{gleiser}
leads to a very significant limit
of $|\eta_\gamma| < 2 \cdot 10^{-4}$.
A even more significant limit on
the  $\eta_\gamma$ parameter
could be inferred from observation of polarized gamma rays
from distant astrophysical sources.
One such observation has been recently reported in the literature:
Ref.~\cite{polarGRB} reports polarized MeV gamma rays in the prompt
emission of the gamma-ray burst GRB021206.
As observed in Ref.~\cite{tedsteck} this would allow to establish
an impressive limit on $\eta_\gamma$ ($\eta_\gamma < 10^{-14}$
or even better). However, the report of
Ref.~\cite{polarGRB}
has been challenged (see {\it e.g.} Ref.~\cite{nopolarGRB}),
and as long as the experimental situation remains unclear
of course these data cannot be used to establish robust experimental
limits.

\section{Closing remarks}
\noindent
With this paper I am hoping to ignite a debate which should lead to
the transition toward a more mature phase of quantum-gravity phenomenology,
in which a key role is played by some reference test theories.
I gave an explicit formulation of
two test theories which could be considered for this role,
and I discussed a few examples of phenomenological analyses
with these two test theories.
The two test theories assume basically the same type of
modification of the dispersion relation, but in my illustrative
examples of phenomenological analyses it emerged that the phenomenology
is in some cases very different.
This exposes the shortcomings of an
approach to the phenomenology
of Planck-scale modified dispersion relations
which had become fashionable
in the recent literature:
there have been several papers claiming to improve limits
on Planck-scale modifications
of the dispersion relation, but the different studies were simply
considering the same type of dispersion relation within
significantly different test theories, or worse the phenomenological
analysis did not even rely on a well-defined test theory.
From outside the quantum-gravity-phenomenology community
these papers were actually perceived as a gradual improvement
in the experimental bounds on the overall idea
of Planck-scale departures from Lorentz symmetry,
to the point that there is now a wide-spread perception
that in general departures from Lorentz symmetry are already
experimentally constrained to be far beyond the Planck-scale.
Instead I showed that two simple and rather natural
test theories
evade automatically some of the possible opportunities
for constraints.

The two test theories on which I focused,
the minimal AEMNS test theory and the minimal
GPMP test theory, could be rather natural starting
points for the two types of intuitions that are
being discussed in the quantum-gravity-phenomenology literature.
The key point is whether we should trust
effective low-energy field theory as the formalism used in the
description of dynamical effects.
The fact that both in the study of noncommutative spacetimes
and in the study of Loop Quantum Gravity, the two quantum
pictures of spacetime that provide the key sources of motivation
for research on Planck-scale modifications of the dispersion relation,
we are really only starting to understand some aspects of kinematics,
but we are still missing any robust result on dynamics,
encourages an approach to phenomenology which
is correspondingly prudent with respect to the description
of dynamics.
Our test theories will be really successful only if they
work well in bridging the gap between experimental data and
our present limited understanding of fundamental
quantum-gravity/quantum-spacetime pictures.
We therefore need a set of test theories reflecting the
different intuitions that are guiding different approaches
to the quantum-gravity problem.
For those who are most concerned about the status of the
description of dynamics in quantum-gravity research,
the assumption of a description of dynamics based on
effective low-energy field theory
appears to be too unreliable,
and the pure-kinematics
minimal AEMNS test theory
may provide a natural starting point for phenomenology.
The fact that the minimal AEMNS test theory does not assume anything
about dynamics of course limits its applicability, but it allows
us to focus (at least in this first stage of investigation)
on the assumption of universality of the modifications of the dispersion
relation. This assumption is in fact fully consistent with the kinematic
structure of the test theory, and may well turn out to be also consistent
with the description of dynamics (when established), if this description
is not field-theory based.

For those who are willing to set aside the concerns about the
description of dynamics,
and go ahead with the effective-field-theory formulation,
the minimal GPMP test theory should provide a valuable starting
point. The fact that this test theory can be compared also to data
involving some aspects of dynamics obviously allows a richer
phenomenology, but the complication of nonuniversality of
the effects must necessarily be accommodated.
In fact the effective-field-theory
formulation automatically requires that the two polarizations of
photons carry opposite-sign modifications of the dispersion relation,
and then a natural criterion (in which the speed-of-light scale preserves
at least its role in the description of
the average speed of randomly-composed particle bursts)
leads to assuming the same sign/helicity correlation also for
all other particles.

As illustrated by the few examples of phenomenological analyses
which I discussed,
the phenomenology work with the minimal  GPMP test theory
is of a rather familiar type. It is a setup that resembles closely
the one of certain nonrenormalizable effective low-energy field theories
used in particle-physics phenomenology (although, as mentioned,
the fine-tuning concerns are more severe).
Instead the
minimal AEMNS test theory, as it is
conceived as a pure-kinematics test theory,
will force us to a type of phenomenology which (to my knowledge)
is new. The abstract idea of a pure-kinematics test theory is well motivated
by the status of our understanding of the relevant quantum-spacetime
frameworks, but, as illustrated by the few examples of phenomenological
analyses which I discussed,
the practical realizations of a pure-kinematics phenomenological analysis
are often confronted with the problem of ``contamination'' by
dynamical effects.
I structured this paper in such a way that, on this crucial point
of the possibility of pure-kinematics analyses,
a certain hierarchy would emerge for the reader.
At the top of this hierarchy
there are some phenomenological analyses
that truly involve pure kinematics, like the time-of-flight
analyses discussed
in subsection IV.A.
Then I considered
the analyses of the photon-absorption threshold and of the
cosmic-ray GZK threshold are examples
of phenomenological analyses that could be used
for pure-kinematics tests in the controlled environment
of a laboratory, where one could control the energies of
the colliding particles, but are subject to a ``conspiracy hypothesis''
in the context of certain applications in astrophysics, where the
incoming-particle energy is known but its potential targets
have energies that spread over a large range.
In those astrophysical applications a conspiracy between the adopted
deformation of kinematics and the unspecified deformation of dynamics
could affect the reliability of the analysis.
And finally there are cases like the one of the synchrotron-radiation
analysis, which even in the controlled laboratory setup could
not be viewed as tests of pure kinematics.
The cases like the synchrotron-radiation analysis, which even in principle
is not structured as a pure-kinematics test, are clearly ill-suited for
the analysis of a pure-kinematics test theory.
More subject to debate is the handling of the analyses which are
subject to a ``conspiracy hypothesis'': while I am advocating a
prudent conservative approach to the derivation of experimental
bounds in this phenomenology, I have argued that, at least for
situations like the one of the AEMNS test theory with $n=1$, $\eta \sim -1$
in the analysis of the GZK threshold, one should sometimes
confidently dismiss the relevant ``conspiracy hypothesis'', which
(as stressed in subsection IV.D) would
require a truly implausible role of a background of multi-TeV
photons.

The adoption of commonly agreed criteria on how to
handle these experimental bounds valid ``up to conspiracies''
would be an important asset for the new phase of
quantum-gravity phenomenology which I am proposing.



\end{document}